 \documentclass[twocolumn]{aastex62}
 
\usepackage{subfigure}
\usepackage{commath}
\usepackage{amsmath}
\usepackage{rotating}
\usepackage{lipsum}
\usepackage{booktabs}
\usepackage{multirow}
\usepackage{floatrow}
\hypersetup{linkcolor=red,citecolor=blue,filecolor=cyan,urlcolor=magenta}

\graphicspath{{./}{figures/}}

\submitjournal{ApJ}

\shorttitle{Alfv\'{e}nic Shock-sheath}
\shortauthors{Raghav \& Shaikh}

\begin{document}

\title{\large{\textbf{The first observable \textit{in-situ} evidence of Alfv\'{e}nic turbulence in shock-sheath}}}

\author[0000-0002-4704-6706]{Anil N. Raghav}
\affiliation{University Department of Physics, University of Mumbai, Vidyanagari, Santacruz (E), Mumbai-400098, India \\
}

\correspondingauthor{Anil Raghav}
\email{raghavanil1984@gmail.com }

\author[0000-0002-9206-6327]{Zubair I. Shaikh}
\affiliation{Indian Institute of Geomagnetism (IIG), New Panvel, Navi Mumbai-410218, India.} 

\correspondingauthor{Zubair Shaikh}
\email{zubairshaikh584@gmail.com}

\begin{abstract}
	
The dynamic evolution of coronal mass ejection (CME) in interplanetary space generates highly turbulent, compressed and heated shock-sheath.
This region furnishes a unique environment to study the turbulent fluctuations at the small scales and serve an opportunity for unfolding the physical mechanisms by which the turbulence is dissipated and plasma is heated. How does the turbulence in the magnetized plasma controls the energy transport process in space and astrophysical plasmas is an attractive and challenging open problem of 21st century. The literature discuss three types of incompressible magnetohydrodynamics (MHD) shocks as the magnetosonic (fast), Alfv\'{e}nic (intermediate), and sonic (slow). The magnetosonic shock is most common in the interplanetary medium.
However, Alfv\'{e}nic shocks have not been identified till date in interplanetary space. In fact, the questions were raised on their existence based on the theoretical ground.
Here, we demonstrate the first observable in-situ evidence of Alfv\'{e}nic turbulent shock-sheath at 1 AU. The study has strong implications in the domain of an interplanetary space plasma, its interaction with planetary plasma and astrophysical plasma.

\end{abstract}




\section{Introduction} 
A coronal mass ejection (CME) is a huge cloud of solar plasma ( mass $\sim 3.2 \times 10^{14}~g$, kinetic energy $\sim 2.0 \times 10^{29}~erg$) submersed in magnetic field lines that are blown away from the Sun which propagates and expands into the interplanetary medium. Their studies are of paramount importance in view of their natural hazardous effects on humans and the technology in space and ground.  \citep{schrijver2010heliophysics,palmer1978bidirectional,low1996solar,gopalswamy2005type,schwenn2006space,moldwin2008introduction,vourlidas2010comprehensive}.
The propagation speed of CMEs is often higher than the ambient solar wind which causes the formation of fast, collision-less shocks ahead of CMEs.These shocks cause heating and compression of the upstream (anti-sunward side) slow solar wind plasma, forming turbulent sheaths between the shocks and the leading edge of the CMEs  \citep{sonett1963distant,kennel1985quarter,papadopoulos1985microinstabilities,tsurutani1985acceleration,echer2011interplanetary,oliveira2014impact,oliveira2015impact,lugaz2016factors,kilpua2017coronal}. 
The shock and sheath are responsible mostly for (i) acceleration  of  solar  energetic  particles  \citep{tsurutani1985acceleration,manchester2005coronal}, (ii) significant geomagnetic activity \citep{tsurutani1988origin}, (iii) Forbush decrease phenomena \citep{raghav2014quantitative,raghav2017forbush,shaikh2017presence,bhaskar2016role}, and 
(iv) Auroral lightning \citep{baker2016resource} etc. Besides this, the shock initiates a magnetosonic wave in the magnetosphere and associated electric field accelerates electrons to MeV energies \citep{foster2015shock,kanekal2016prompt}.
Recently, the loss of electron flux from the radiation belt has been observed during the shock-sheath encounter with Earth's magnetosphere \citep{hietala2014depleting, kilpua2015unraveling,kilpua2017coronal}. This may be caused due to an increase in ultra-low frequency (ULF) wave power and dynamic pressure which is further responsible for pitch angle scattering and radial diffusion of the electron flux.
The precipitated high energy electron flux from radiation belt is used as a key parameter in climate models and in the understanding of atmospheric chemistry and associated climatological effects \citep{andersson2014missing,verronen2011first,mironova2015energetic}. In addition to this, the other planets and their atmospheres are highly affected by the shock-sheath of CME, for example, in case of Mars loss of the ions flux ($>~9~amu$) is observed which might be caused by its high dynamic pressure \citep{jakosky2015maven}.

CME induced shock-sheath provides unique opportunity to investigate the nature of plasma turbulence, plasma energy/fluctuation dissipation, and plasma heating process. 
The plasma turbulence demonstrates the features such as Alfv\'{e}n waves, Whistler waves,  ion cyclotron waves, or ion Bernstein waves etc \citep{schekochihin2009astrophysical,salem2012identification,shaikh2010inhomogeneous,krishan2004magnetic,gary2009short,he2011possible,sahraoui2012new}. In fact, sometimes plasma fluctuations do not exhibit any wave-like configuration at all but resemble nonlinear structures such as current sheets \citep{osman2010evidence,sundkvist2007dissipation}.
Actually various  studies related to the nature of  turbulence and generation of wave in shock-sheath region have been reported in recent past.  \citet{liu2006plasma} observed the mirror mode wave within the shock-sheath region . \citet{kilpua2013magnetic}
observed that the power of ultra-low frequency fluctuation (in the dynamic pressure and magnetic field) peaks close to the shock-front and sheath-magnetic cloud boundary. Furthermore,  large amplitude magnetic field fluctuations, as well as intense irregular ULF fluctuations and regular high-frequency wave activity is also observed in the downstream of CME shocks \citep{kataoka2005downstream,kajdivc2012waves,goncharov2014upstream}. Moreover, Whistler waves associated with weak interplanetary shocks are also observed \citep{ramirez2012whistler}. However, Alfven wave within the shock-sheath of CME was not observed yet.
In fact, literature indicates theoretical debate on their existence  \citep{wu1987mhd,kunkel1966plasma,taniuti1964non}. To the best of our knowledge, here we demonstrate the first {in-situ} observable evidence of Alfv\'{e}n wave within the shock-sheath region of CME.

\section{Event details $\&$ Alfv\'{e}n wave confirmation} 

\begin{figure}
	\begin{center}
		
		\includegraphics[width=\textwidth]{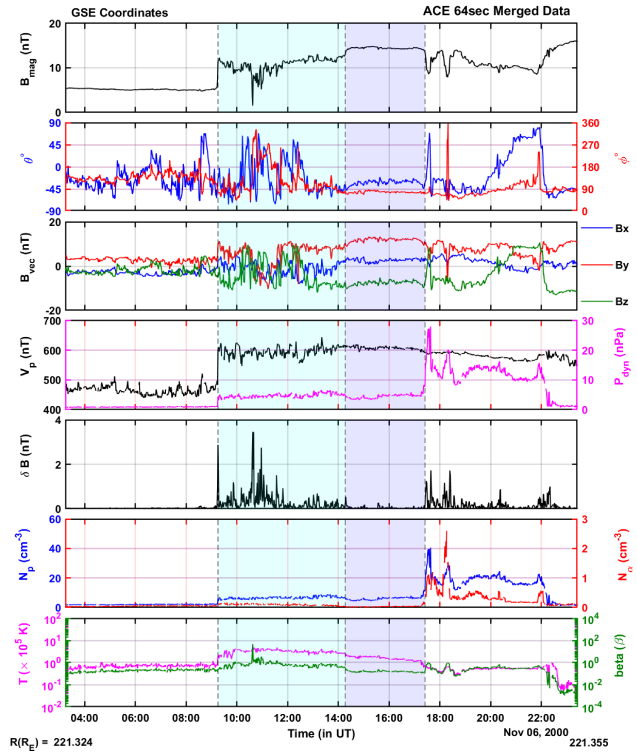}
		\caption{The CME observed by the ACE spacecraft on $06^{th} ~ November, ~ 2000$ with time cadence of 64 sec. The top to bottom panels represents different interplanetary parameters such as: total interplanetary field strength IMF $B_{mag}$, azimuth ($\phi$) $\&$ elevation ($\theta$) angle,  IMF vectors i.e. $B_{vec}$, Plasma velocity ($V_p$) $\&$ Plasma dynamics pressure ($P_{dyn}$), $\delta B$,  Proton $\&$ Alpha density ($N_p$, $N_{\alpha}$), and  Temperature ($T_p$) $\&$ plasma beta ($\beta$) respectively. The shaded regions represents the shock-sheath of CME (cyan) and its magnetic cloud (purple). All observations are in GSE coordinate system. 
		}
		\label{fig:IP}
	\end{center}
\end{figure}

The shock-sheath under investigation is engendered by a CME which crossed the WIND and ACE spacecrafts on $06^{th}$ November $2000$. Figure \ref{fig:IP} demonstrates the temporal variations of various in-situ  plasma parameters and the  interplanetary magnetic field (IMF) measured by the ACE spacecraft (The Wind spacecraft measurements are also studied, however not presented here). The commencement of the shock at spacecraft is identified as a sudden enhancement in the $B_{mag}$, $V_p$, $N_p$, $T_p$, and $\beta$ and it is indicated by the first vertical black dashed line. The confirmation of the shock and its properties are given at \url{https://www.cfa.harvard.edu/shocks/ac_master_data/00076/ac_master_00076.html}. The shock is followed by large fluctuations in IMF (See $\delta B$) with enhanced magnetic field strength; high $N_p$, $T_p$ \& $P_{dyn}$  which is manifested as a shock-sheath region. The second shaded region shows the least fluctuations in $B_{mag}$ and its components, the slow variation in $\theta$ and $\phi$, the slow steady trend in $V_p$ and low $\beta$. This indicates the presence of a CME magnetic cloud region \citep{zurbuchen2006situ}. 

\begin{figure}
	\begin{center}
		\includegraphics[width=\textwidth]{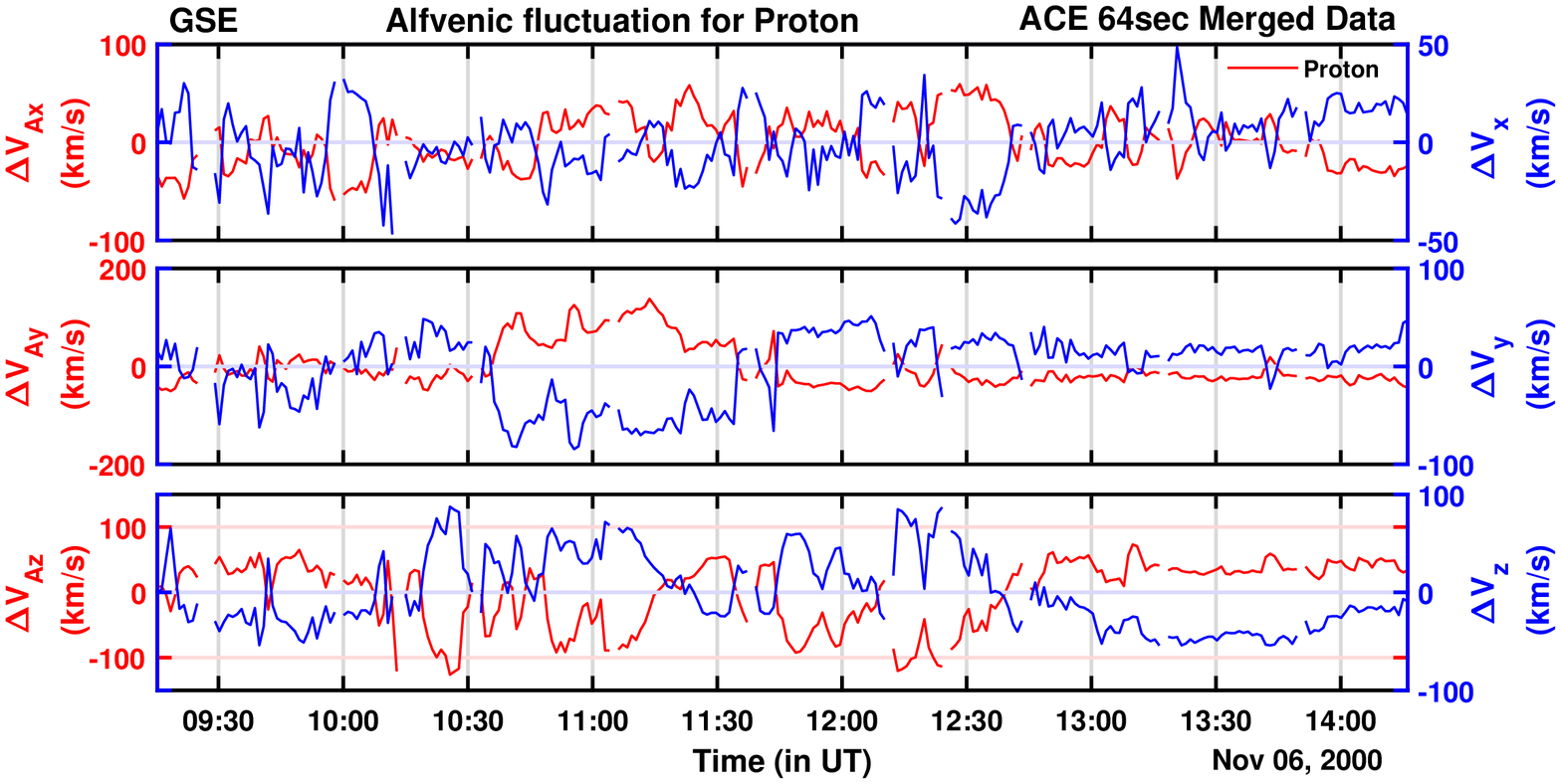}
		\caption{The  temporal variation of Alfv\'{e}n velocity fluctuation vector $\Delta V_A$ (red)  and that of the proton flow velocity fluctuation vector $\Delta V$ (blue).
		}
	\end{center}
	\label{fig:Alfven_velocity}
\end{figure}

\begin{figure}
	\includegraphics[width=\textwidth]{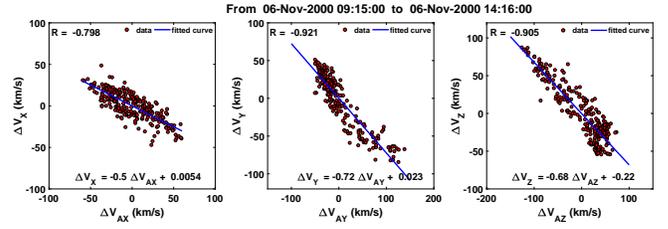}
	\caption{The correlation and regression analysis between the respective $\Delta V$ and $\Delta V_{A}$ components. The scattered black circle with filled red color represent observations from ACE spacecraft with time cadence of $64~s$. The $R$ is the  coefficient of correlation. The equation in each panel indicate the linear fit relation between respective components of $\Delta V$ and $\Delta V_{A}$.}
	\label{fig:Walen_p}
\end{figure}

A typical Wal\'{e}n test is employed to confirm the presence of the Alfv\'{e}n wave in the solar wind. The Wal\'{e}n relation is described as \citep{walen1944theory,hudson1971rotational,yang2013alfven,yang2016observational,raghav2018first,raghav2018does}:
\begin{equation}
V_A = \pm A \frac{B}{\sqrt{\mu_0 \rho}}
\end{equation}
where B is a magnetic field, A is the anisotropy parameter, and $\rho$ is proton mass density. Normally, $A = \pm 1$ for negligible thermal anisotropic plasma. We can obtain $\Delta B$ by subtracting an average value of $B$ from each measured value. Hence, the fluctuation in Alfv\'{e}n velocity is given as: 
\begin{equation}
\Delta V_A = \pm A \frac{\Delta B}{\sqrt{\mu_0 \rho}}
\end{equation}
Further, we calculate $\Delta V$ by subtracting averaged proton flow velocity from measured values. 
Figure \ref{fig:Alfven_velocity} represents the comparisons of $x$, $y$, and $z$ components of $\Delta V_A$ and $\Delta V$ respectively. It clearly demonstrates correlated variations between their respective components within the shock-sheath region and indicates the possible existence of Alfv\'{e}n wave. Figure \ref{fig:Walen_p} represents their correlation and regression analysis. The Pearson correlation coefficients (R) of each $x$, $y$, and $z$ components are $-0.798$, $-0.921$, and $-0.905$ and the corresponding regression slopes are $-0.50$, $-0.72$, and $-0.68$ respectively. The strong negative correlation confirms the presence of the sun-ward Alfv\'{e}n wave in the shock-sheath region of the CME \citep{gosling2010torsional,marubashi2010torsional,zhang2014alfven}.

\section{Analysis $\&$ Discussion}
\begin{figure}
	\includegraphics[width=\textwidth]{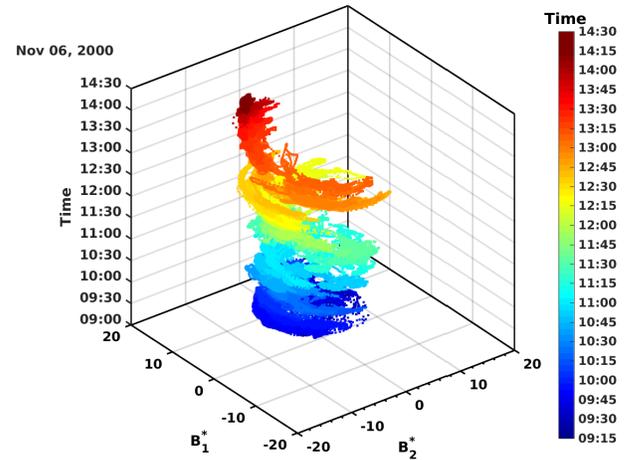}
	\caption{The arc polarization of Alfv\'{e}n wave. Here, $B_1$ and $B_2$ are magnetic field vector corresponding to maximum and intermediate eigenvalue. The color-bar represents the time evolution of the wave.}
	\label{fig:hodo}
\end{figure}

We performed minimum and maximum variance analysis (MVA) of the magnetic field data for the shock-sheath region to study the features of the Alfv\'{e}n wave. The $B_1^*$, $B_2^*$, and $B_3^*$ are the magnetic field vectors after MVA analysis corresponding to maximum, intermediate, and  minimum variance direction respectively. Figure \ref{fig:hodo} shows the time evolution of the Alfv\'{e}n wave which demonstrates that the wave starts at one point, completes the half circle and then returns to the starting position. This behavior of the wave suggests a feature of arc polarization \citep{tsurutani1995interplanetary,riley1995alfvenic,riley1996properties}. Figure \ref{fig:hodo} depicts that the wave does not have an even rate of wave phase rotation with time indicating phase-steepened phenomena. In each arc, there is $~180^\circ$ phase rotation associated with the initial to an intermediate portion of the wave and the trailing portion of the wave carries $~180^\circ$ of phase rotation (half circle). Therefore, the wave period has doubled and indicates the period doubling phenomenon. These characteristics of Alfv\'{e}n wave are discussed in detail by \citet{tsurutani2018review}  and references therein.

\begin{figure}
	\includegraphics[width= \textwidth]{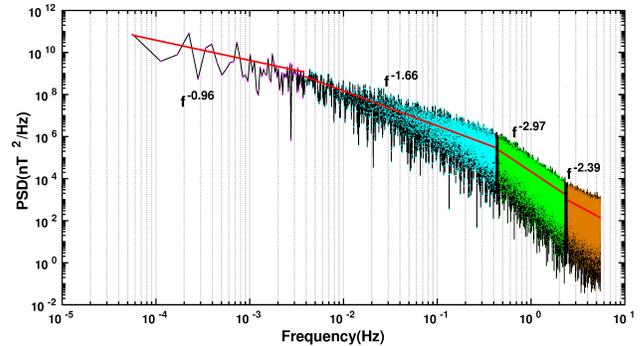}
	\caption{Power spectra of magnetic fluctuations along the $B_x$ direction in GSE coordinate (black color) as a function of frequency in the spacecraft frame as measured by Wind on 2000 November 06, from 09:15 to 14:16 UTC, computed with FFT (black) algorithms. The straight red lines show power-law fits. We have used WINDS high resolution (11 Hz) IMF data. Different colored region represents the data points used for fitting.}
	\label{fig:PSD}
\end{figure}

We performed power spectral density (PSD) analysis to study the characterization of the multi-scale nature of shock-sheath turbulences/fluctuations. Figure \ref{fig:PSD} represents the spectral output for the $B_x$ component of the IMF (other components of IMF are also studied and they shows similar behavior, but not shown here).  The lower frequency is characterized by a flicker noise $f^{-0.96}$ spectrum, which may be due to the residual uncorrelated coronal structures \citep{matthaeus2007density,telloni2009statistics}. This implies the equal energy per interval, which is independent of $f$, in $6\times10^{-5}Hz-4\times10^{-3}Hz$ range of frequencies. The common feature of the incompressible MHD plasma turbulence i.e. Kolmogorov-like turbulence is observed in the intermediate frequency range which is consistent with $\sim f^{-1.66}$ \citep{bruno2005solar}. We believed that the entire turbulent interactions within these regimes are governed by the Alfv\'{e}nic cascade. This region follows and extends up to the ion gyro-frequency. 

Figure \ref{fig:PSD} shows the spectral break at $\sim 0.42~Hz$, and $\sim 2.2~Hz$ in high-frequency region.
We estimate cyclotron frequencies for proton and alpha particle for the observed magnetic field of $5-12~nT$ range, which turn out to be $0.47 ~Hz - 1.10~Hz$ for proton and $0.24 ~Hz - 0.55~Hz$ for alpha. The various studies reveal that at length-scales beyond the  MHD regime (i.e. length scale $<$ ion gyro-radius and temporal scale $>$ ion cyclotron frequency), the power spectrum shows spectral break which halts the Alfv\'{e}nic cascade. Moreover, the presence of the modes in this region evolves on the time-scales associated with dispersive kinetic Alfv\'{e}nic fluctuations \citep{leamon1999dynamics,alexandrova2008small,bale2005measurement,sahraoui2009evidence,goldstein1994properties,shaikh20093d}. However, \citet{perri2010does} suggest that the spectral break in the solar wind is independent of the distance from the Sun, and that of both the ion-cyclotron frequency and the proton gyro-radius. Therefore, it is also possible that the observed high-frequency break in our study is caused by a combination of different physical processes as a result of high compression within the shock-sheath region. The other possible mechanism for the spectral break may result from energy transfer processes related to; 1) kinetic Alfv\'{e}n wave (KAW) \citep{hasegawa1976parametric}, 2) electromagnetic ion-cyclotron-Alfv\'{e}n (EMICA) waves \citep{wu2007proton,gary2008cascade}, or the fluctuation associated with the  Hall  magnetohydrodynamics (HMHD) plasma model \citep{alexandrova2008small,alexandrova2007solar}.

At higher frequencies, the spectrum of the magnetic field fluctuations has power-law dependence as $\sim f^{-\alpha}$, where, the value of $\alpha$ may range from 2 to 4. The average value of the $\alpha $ is close to 7/3 \citep{leamon1998observational,smith2006dependence,alexandrova2008small}. In our study, it is about $\sim {-2.97}$ and $\sim {-2.39}$. These higher frequency part of the spectrum may be associated either with a dissipative range \citep{leamon1998observational,smith2006dependence} or with a different turbulent energy cascade caused by dispersive effects \citep{alexandrova2008small,stasiewicz2000small,li2016new,sahraoui2006anisotropic,sahraoui2009evidence,galtier2007multiscale}. \citet{stawicki2001solar}  proposed that suppression of the Alfv\'{e}nic fluctuations are due to  the ion cyclotron damping at intermediate wave frequency (wavenumber), hence, the observed power spectra  are weakly damped dispersive magnetosonic  and/or whistler waves (unlike Alfven waves). Presence of the whistler wave mode in high-frequency regime was proposed by the \citet{beinroth1981properties}. \citet{goldstein1994properties} found out the existence of multi-scale waves (Alfvenic, whistlers and cyclotron  waves) with a single polarization in the dissipation regime of the spectrum. Observation of the obliquely propagating KAWs (in the $\omega < \omega_{ci}$ regime or Alfvenic regime) puts a question about the spectral breakpoint due to  damping  of ion cyclotron waves \citep{howes2008kinetic}. The Kinetic \citep{howes2008kinetic} and Fluid  \citep{shaikh20093d} simulations show that the ion inertial length-scale is comparable to that of  the spectral breakpoint near  the characteristic turbulent length-scales. For the length-scales larger than the  ion inertial length-scales, the simulations demonstrate  Kolmogorov-like spectra. Moreover, for smaller ion inertial length-scales, they observed  steeper spectrum that is close to $f^{-7/3}$.

\section{Results $\&$ Implications}

In this study, we find a good correlation between the fluctuations of the magnetic field  and proton velocity vectors using the Wal\'{e}n test. This implies that the fluid velocity and magnetic field are fluctuating together and propagate along the direction of the magnetic tension force. The observed feature confirms the presence of Alfv\'{e}n wave in shock-sheath. The Sunward Alfv\'{e}n waves are very rare as compared to anti-Sunward Alfv\'{e}n waves in the solar wind
\citep{belcher1969large,daily1973alfven,burlaga1976microscale,riley1996properties,denskat1982statistical,yang2016observational}. They are associated with events such as magnetic reconnection exhausts and/or back-streaming ions from reverse shocks and are expected with increasing heliocentric distance \citep{belcher1971large,roberts1987origin,bavassano1989large,gosling2009one,gosling2011pulsed}. The Alfv\'{e}n waves may be generated by (i) the steepening of a magnetosonic wave which forms the shock at the leading edge of the magnetic cloud \citep{tsurutani1988origin,tsurutani2011review}, (ii) velocity shear instabilities \citep{bavassano1978local,coleman1968turbulence,roberts1987origin,roberts1992velocity}, (iii) the oblique firehose instability \citep{matteini2007evolution,matteini2006parallel,hellinger2008oblique}, (iv) kinetic instabilities
associated with the solar wind proton heat flux \citep{goldstein2000observed,matteini2013signatures,marsch1987observational},  and (v) the interaction of multiple CMEs  \citep{raghav2018first} etc. Moreover, the Alfv\'{e}n waves are commonly observed in interplanetary space \citep{hellinger2008oblique} but it would be difficult for them to get into the magnetic cloud.

The MVA technique indicates Alfv\'{e}n wave characteristics such as, arc polarization, phase steepening and period doubling. The  Alfv\'{e}n waves leads to non-linear interactions \citep{dobrowolny1980fully} which are crucial for the dynamical evolution of a Kolmogorov-like MHD spectrum \citep{bruno2013solar}. The PSD analysis depicts turbulent nature of the shock-sheath. Thus, we  observed  the continued cascade of energy from large scales to smaller scales of wavelengths and eventually to such small scales that the plasma no longer behaves like a fluid due to change in velocity and magnetic field fluctuations. At this scale, the particle distribution is affected by the magnetic field which may lead to plasma heating. The heating process might happen through either resonant interactions or stochastic jumps. We opine that the plasma heating in shock-sheath could be associated with an above-discussed process. The various simulation studies and spacecraft data analysis in solar wind reveals that the turbulent energy transfers essentially through different regions in the k-space, i.e. $k^{-1}$, $k^{-5/3}$, and $k^{-7/3}$. Our results are clearly consistent with the reported studies and confirm that the shock-sheath is dominated by Alfv\'{e}nic turbulence.

However, several open questions need to be addressed in view of turbulent nature in highly compressed and heated shock-sheath such as, (i) What is the origin of turbulent cascade in shock-sheath? Is it the coronal plasma or local driving?; (ii) How does the cascade modify the shock-sheath plasma?; (iii) How do the turbulent fluctuations get dissipated into heat?; (iv) What is more important for energy dissipation, non-linear turbulent heating or resonant wave-particle interactions?; (v) Can shock-sheath turbulence be parameterized and included in heliospheric models for space weather prediction?

Recently, the presence of the Alfv\'{e}n wave has been seen in a magnetic cloud of CME \citep{raghav2018first}. It is manifested that the Alfv\'{e}nic oscillations in a magnetic cloud of CME may cause the internal magnetic reconnection and/or thermal anisotropy in plasma distribution which leads to the disruption of the stable magnetic structure of the CME \citep{raghav2018does}. Their presence in the magnetic cloud of CME also controls the recovery phase of the geomagnetic storms \citep{raghav2018torsional}. In the introduction section of the article, we emphasize that the shock-sheath of CME not only affects the interplanetary plasma characteristics but also affects the dynamics of magnetosphere, ionosphere, radiation-belt and upper atmosphere of the Earth. In fact, it affects the other planetary exospheres as well. Therefore how the typical configuration such as Alfv\'{e}n fluctuations embedded shock-sheath influence the overall solar-terrestrial plasma will be intriguing and may activate the possible direction of future studies. One can also expect similar features of shock-sheath in interstellar medium as well e.g. supernovae shocks and associated sheaths.

\section*{Acknowledgement}
We are thankful to ACE and WIND Spacecraft data providers for  making interplanetary data available. AR is also thakful to
Gauri Datar.

\acknowledgments

\vspace{2cm}
\newpage

\bibliographystyle{aasjournal}
\vspace{2cm}

\begin{thebibliography}{}
	\expandafter\ifx\csname natexlab\endcsname\relax\def\natexlab#1{#1}\fi
	\providecommand{\url}[1]{\href{#1}{#1}}
	
	\bibitem[{Alexandrova {et~al.}(2007)Alexandrova, Carbone, Veltri, \&
		Sorriso-Valvo}]{alexandrova2007solar}
	Alexandrova, O., Carbone, V., Veltri, P., \& Sorriso-Valvo, L. 2007, Planetary
	and Space Science, 55, 2224
	
	\bibitem[{Alexandrova {et~al.}(2008)Alexandrova, Carbone, Veltri, \&
		Sorriso-Valvo}]{alexandrova2008small}
	---. 2008, The Astrophysical Journal, 674, 1153
	
	\bibitem[{Andersson {et~al.}(2014)Andersson, Verronen, Rodger, Clilverd, \&
		Sepp{\"a}l{\"a}}]{andersson2014missing}
	Andersson, M., Verronen, P., Rodger, C., Clilverd, M., \& Sepp{\"a}l{\"a}, A.
	2014, Nature communications, 5, 5197
	
	\bibitem[{Baker \& Lanzerotti(2016)}]{baker2016resource}
	Baker, D.~N., \& Lanzerotti, L.~J. 2016, American Journal of Physics, 84, 166
	
	\bibitem[{Bale {et~al.}(2005)Bale, Kellogg, Mozer, Horbury, \&
		Reme}]{bale2005measurement}
	Bale, S., Kellogg, P., Mozer, F., Horbury, T., \& Reme, H. 2005, Physical
	Review Letters, 94, 215002
	
	\bibitem[{Bavassano \& Bruno(1989)}]{bavassano1989large}
	Bavassano, B., \& Bruno, R. 1989, Journal of Geophysical Research: Space
	Physics, 94, 168
	
	\bibitem[{Bavassano {et~al.}(1978)Bavassano, Dobrowolny, \&
		Moreno}]{bavassano1978local}
	Bavassano, B., Dobrowolny, M., \& Moreno, G. 1978, Solar Physics, 57, 445
	
	\bibitem[{Beinroth \& Neubauer(1981)}]{beinroth1981properties}
	Beinroth, H., \& Neubauer, F. 1981, Journal of Geophysical Research: Space
	Physics, 86, 7755
	
	\bibitem[{Belcher \& Davis(1971)}]{belcher1971large}
	Belcher, J., \& Davis, L. 1971, Journal of Geophysical Research, 76, 3534
	
	\bibitem[{Belcher {et~al.}(1969)Belcher, Davis, \& Smith}]{belcher1969large}
	Belcher, J., Davis, L., \& Smith, E. 1969, Journal of Geophysical Research, 74,
	2302
	
	\bibitem[{Bhaskar {et~al.}(2016)Bhaskar, Vichare, Arunbabu, \&
		Raghav}]{bhaskar2016role}
	Bhaskar, A., Vichare, G., Arunbabu, K., \& Raghav, A. 2016, Astrophysics and
	Space Science, 361, 242
	
	\bibitem[{Bruno \& Carbone(2005)}]{bruno2005solar}
	Bruno, R., \& Carbone, V. 2005, Living Reviews in Solar Physics, 2, 4
	
	\bibitem[{Bruno \& Carbone(2013)}]{bruno2013solar}
	---. 2013, Living Reviews in Solar Physics, 10, 2
	
	\bibitem[{Burlaga \& Turner(1976)}]{burlaga1976microscale}
	Burlaga, L., \& Turner, J. 1976, Journal of Geophysical Research, 81, 73
	
	\bibitem[{Coleman~Jr(1968)}]{coleman1968turbulence}
	Coleman~Jr, P.~J. 1968, The Astrophysical Journal, 153, 371
	
	\bibitem[{Daily(1973)}]{daily1973alfven}
	Daily, W.~D. 1973, Journal of Geophysical Research, 78, 2043
	
	\bibitem[{Denskat \& Neubauer(1982)}]{denskat1982statistical}
	Denskat, K., \& Neubauer, F. 1982, Journal of Geophysical Research: Space
	Physics, 87, 2215
	
	\bibitem[{Dobrowolny {et~al.}(1980)Dobrowolny, Mangeney, \&
		Veltri}]{dobrowolny1980fully}
	Dobrowolny, M., Mangeney, A., \& Veltri, P. 1980, Physical Review Letters, 45,
	144
	
	\bibitem[{Echer {et~al.}(2011)Echer, Tsurutani, Guarnieri, \&
		Kozyra}]{echer2011interplanetary}
	Echer, E., Tsurutani, B., Guarnieri, F., \& Kozyra, J. 2011, Journal of
	Atmospheric and Solar-Terrestrial Physics, 73, 1330
	
	\bibitem[{Foster {et~al.}(2015)Foster, Wygant, Hudson, Boyd, Baker, Erickson,
		\& Spence}]{foster2015shock}
	Foster, J., Wygant, J., Hudson, M., {et~al.} 2015, Journal of Geophysical
	Research: Space Physics, 120, 1661
	
	\bibitem[{Galtier \& Buchlin(2007)}]{galtier2007multiscale}
	Galtier, S., \& Buchlin, E. 2007, The Astrophysical Journal, 656, 560
	
	\bibitem[{Gary {et~al.}(2008)Gary, Saito, \& Li}]{gary2008cascade}
	Gary, S.~P., Saito, S., \& Li, H. 2008, Geophysical Research Letters, 35
	
	\bibitem[{Gary \& Smith(2009)}]{gary2009short}
	Gary, S.~P., \& Smith, C.~W. 2009, Journal of Geophysical Research: Space
	Physics, 114
	
	\bibitem[{Goldstein {et~al.}(2000)Goldstein, Neugebauer, Zhang, \&
		Gary}]{goldstein2000observed}
	Goldstein, B.~E., Neugebauer, M., Zhang, L.~D., \& Gary, S.~P. 2000,
	Geophysical research letters, 27, 53
	
	\bibitem[{Goldstein {et~al.}(1994)Goldstein, Roberts, \&
		Fitch}]{goldstein1994properties}
	Goldstein, M., Roberts, D., \& Fitch, C. 1994, Journal of Geophysical Research:
	Space Physics, 99, 11519
	
	\bibitem[{Goncharov {et~al.}(2014)Goncharov, {\v{S}}afr{\'a}nkov{\'a},
		N{\v{e}}me{\v{c}}ek, P{\v{r}}ech, Pit{\v{n}}a, \&
		Zastenker}]{goncharov2014upstream}
	Goncharov, O., {\v{S}}afr{\'a}nkov{\'a}, J., N{\v{e}}me{\v{c}}ek, Z., {et~al.}
	2014, Geophysical Research Letters, 41, 8100
	
	\bibitem[{Gopalswamy {et~al.}(2005)Gopalswamy, Aguilar-Rodriguez, Yashiro,
		Nunes, Kaiser, \& Howard}]{gopalswamy2005type}
	Gopalswamy, N., Aguilar-Rodriguez, E., Yashiro, S., {et~al.} 2005, Journal of
	Geophysical Research: Space Physics, 110
	
	\bibitem[{Gosling {et~al.}(2009)Gosling, McComas, Roberts, \&
		Skoug}]{gosling2009one}
	Gosling, J., McComas, D., Roberts, D., \& Skoug, R. 2009, The Astrophysical
	Journal Letters, 695, L213
	
	\bibitem[{Gosling {et~al.}(2010)Gosling, Teh, \&
		Eriksson}]{gosling2010torsional}
	Gosling, J., Teh, W.-L., \& Eriksson, S. 2010, The Astrophysical Journal
	Letters, 719, L36
	
	\bibitem[{Gosling {et~al.}(2011)Gosling, Tian, \& Phan}]{gosling2011pulsed}
	Gosling, J., Tian, H., \& Phan, T. 2011, The Astrophysical Journal Letters,
	737, L35
	
	\bibitem[{Hasegawa \& Chen(1976)}]{hasegawa1976parametric}
	Hasegawa, A., \& Chen, L. 1976, Physical Review Letters, 36, 1362
	
	\bibitem[{He {et~al.}(2011)He, Marsch, Tu, Yao, \& Tian}]{he2011possible}
	He, J., Marsch, E., Tu, C., Yao, S., \& Tian, H. 2011, The Astrophysical
	Journal, 731, 85
	
	\bibitem[{Hellinger \& Tr{\'a}vn{\'\i}{\v{c}}ek(2008)}]{hellinger2008oblique}
	Hellinger, P., \& Tr{\'a}vn{\'\i}{\v{c}}ek, P.~M. 2008, Journal of Geophysical
	Research: Space Physics, 113
	
	\bibitem[{Hietala {et~al.}(2014)Hietala, Kilpua, Turner, \&
		Angelopoulos}]{hietala2014depleting}
	Hietala, H., Kilpua, E., Turner, D., \& Angelopoulos, V. 2014, Geophysical
	Research Letters, 41, 2258
	
	\bibitem[{Howes {et~al.}(2008)Howes, Dorland, Cowley, Hammett, Quataert,
		Schekochihin, \& Tatsuno}]{howes2008kinetic}
	Howes, G., Dorland, W., Cowley, S., {et~al.} 2008, Physical Review Letters,
	100, 065004
	
	\bibitem[{Hudson(1971)}]{hudson1971rotational}
	Hudson, P. 1971, Planetary and Space Science, 19, 1693
	
	\bibitem[{Jakosky {et~al.}(2015)Jakosky, Grebowsky, Luhmann, Connerney,
		Eparvier, Ergun, Halekas, Larson, Mahaffy, Mcfadden,
		{et~al.}}]{jakosky2015maven}
	Jakosky, B.~M., Grebowsky, J.~M., Luhmann, J.~G., {et~al.} 2015, Science, 350,
	aad0210
	
	\bibitem[{Kajdi{\v{c}} {et~al.}(2012)Kajdi{\v{c}}, Blanco-Cano,
		Aguilar-Rodriguez, Russell, Jian, \& Luhmann}]{kajdivc2012waves}
	Kajdi{\v{c}}, P., Blanco-Cano, X., Aguilar-Rodriguez, E., {et~al.} 2012,
	Journal of Geophysical Research: Space Physics, 117
	
	\bibitem[{Kanekal {et~al.}(2016)Kanekal, Baker, Fennell, Jones, Schiller,
		Richardson, Li, Turner, Califf, Claudepierre, {et~al.}}]{kanekal2016prompt}
	Kanekal, S., Baker, D., Fennell, J., {et~al.} 2016, Journal of Geophysical
	Research: Space Physics, 121, 7622
	
	\bibitem[{Kataoka {et~al.}(2005)Kataoka, Watari, Shimada, Shimazu, \&
		Marubashi}]{kataoka2005downstream}
	Kataoka, R., Watari, S., Shimada, N., Shimazu, H., \& Marubashi, K. 2005,
	Geophysical research letters, 32
	
	\bibitem[{Kennel {et~al.}(1985)Kennel, Edmiston, \& Hada}]{kennel1985quarter}
	Kennel, C., Edmiston, J., \& Hada, T. 1985, Collisionless shocks in the
	heliosphere: A tutorial review, 34, 1
	
	\bibitem[{Kilpua {et~al.}(2013)Kilpua, Hietala, Koskinen, Fontaine, Turc,
		{et~al.}}]{kilpua2013magnetic}
	Kilpua, E., Hietala, H., Koskinen, H., {et~al.} 2013, in Annales Geophysicae
	
	\bibitem[{Kilpua {et~al.}(2017)Kilpua, Koskinen, \&
		Pulkkinen}]{kilpua2017coronal}
	Kilpua, E., Koskinen, H.~E., \& Pulkkinen, T.~I. 2017, Living Reviews in Solar
	Physics, 14, 5
	
	\bibitem[{Kilpua {et~al.}(2015)Kilpua, Hietala, Turner, Koskinen, Pulkkinen,
		Rodriguez, Reeves, Claudepierre, \& Spence}]{kilpua2015unraveling}
	Kilpua, E., Hietala, H., Turner, D., {et~al.} 2015, Geophysical Research
	Letters, 42, 3076
	
	\bibitem[{Krishan \& Mahajan(2004)}]{krishan2004magnetic}
	Krishan, V., \& Mahajan, S. 2004, Journal of Geophysical Research: Space
	Physics, 109
	
	\bibitem[{Kunkel(1966)}]{kunkel1966plasma}
	Kunkel, W.~B. 1966, Plasma physics in theory and application (McGraw-Hill)
	
	\bibitem[{Leamon {et~al.}(1999)Leamon, Ness, Smith, \&
		Wong}]{leamon1999dynamics}
	Leamon, R.~J., Ness, N.~F., Smith, C.~W., \& Wong, H.~K. 1999in , AIP, 469--472
	
	\bibitem[{Leamon {et~al.}(1998)Leamon, Smith, Ness, Matthaeus, \&
		Wong}]{leamon1998observational}
	Leamon, R.~J., Smith, C.~W., Ness, N.~F., Matthaeus, W.~H., \& Wong, H.~K.
	1998, Journal of Geophysical Research: Space Physics, 103, 4775
	
	\bibitem[{Li {et~al.}(2016)Li, Wang, Chao, \& Hsieh}]{li2016new}
	Li, H., Wang, C., Chao, J., \& Hsieh, W. 2016, Journal of Geophysical Research:
	Space Physics, 121, 42
	
	\bibitem[{Liu {et~al.}(2006)Liu, Richardson, Belcher, Kasper, \&
		Skoug}]{liu2006plasma}
	Liu, Y., Richardson, J., Belcher, J., Kasper, J., \& Skoug, R. 2006, Journal of
	Geophysical Research: Space Physics, 111
	
	\bibitem[{Low(1996)}]{low1996solar}
	Low, B. 1996, Solar Physics, 167, 217
	
	\bibitem[{Lugaz {et~al.}(2016)Lugaz, Farrugia, Winslow, Al-Haddad, Kilpua, \&
		Riley}]{lugaz2016factors}
	Lugaz, N., Farrugia, C., Winslow, R., {et~al.} 2016, Journal of Geophysical
	Research: Space Physics, 121
	
	\bibitem[{Manchester~IV {et~al.}(2005)Manchester~IV, Gombosi, De~Zeeuw,
		Sokolov, Roussev, Powell, K{\'o}ta, T{\'o}th, \&
		Zurbuchen}]{manchester2005coronal}
	Manchester~IV, W., Gombosi, T., De~Zeeuw, D., {et~al.} 2005, The Astrophysical
	Journal, 622, 1225
	
	\bibitem[{Marsch \& Livi(1987)}]{marsch1987observational}
	Marsch, E., \& Livi, S. 1987, Journal of Geophysical Research: Space Physics,
	92, 7263
	
	\bibitem[{Marubashi {et~al.}(2010)Marubashi, Cho, \&
		Park}]{marubashi2010torsional}
	Marubashi, K., Cho, K.-S., \& Park, Y.-D. 2010in , AIP, 240--244
	
	\bibitem[{Matteini {et~al.}(2013)Matteini, Hellinger, Goldstein, Landi, Velli,
		\& Neugebauer}]{matteini2013signatures}
	Matteini, L., Hellinger, P., Goldstein, B.~E., {et~al.} 2013, Journal of
	Geophysical Research: Space Physics, 118, 2771
	
	\bibitem[{Matteini {et~al.}(2007)Matteini, Landi, Hellinger, Pantellini,
		Maksimovic, Velli, Goldstein, \& Marsch}]{matteini2007evolution}
	Matteini, L., Landi, S., Hellinger, P., {et~al.} 2007, Geophysical Research
	Letters, 34
	
	\bibitem[{Matteini {et~al.}(2006)Matteini, Landi, Hellinger, \&
		Velli}]{matteini2006parallel}
	Matteini, L., Landi, S., Hellinger, P., \& Velli, M. 2006, Journal of
	Geophysical Research: Space Physics, 111
	
	\bibitem[{Matthaeus {et~al.}(2007)Matthaeus, Breech, Dmitruk, Bemporad,
		Poletto, Velli, \& Romoli}]{matthaeus2007density}
	Matthaeus, W., Breech, B., Dmitruk, P., {et~al.} 2007, The Astrophysical
	Journal Letters, 657, L121
	
	\bibitem[{Mironova {et~al.}(2015)Mironova, Aplin, Arnold, Bazilevskaya,
		Harrison, Krivolutsky, Nicoll, Rozanov, Turunen, \&
		Usoskin}]{mironova2015energetic}
	Mironova, I.~A., Aplin, K.~L., Arnold, F., {et~al.} 2015, Space science
	reviews, 194, 1
	
	\bibitem[{Moldwin(2008)}]{moldwin2008introduction}
	Moldwin, M. 2008, An introduction to space weather (Cambridge University Press)
	
	\bibitem[{Oliveira \& Raeder(2014)}]{oliveira2014impact}
	Oliveira, D.~M., \& Raeder, J. 2014, Journal of Geophysical Research: Space
	Physics, 119, 8188
	
	\bibitem[{Oliveira \& Raeder(2015)}]{oliveira2015impact}
	---. 2015, Journal of Geophysical Research: Space Physics, 120, 4313
	
	\bibitem[{Osman {et~al.}(2010)Osman, Matthaeus, Greco, \&
		Servidio}]{osman2010evidence}
	Osman, K., Matthaeus, W., Greco, A., \& Servidio, S. 2010, The Astrophysical
	Journal Letters, 727, L11
	
	\bibitem[{Palmer {et~al.}(1978)Palmer, Allum, \&
		Singer}]{palmer1978bidirectional}
	Palmer, I., Allum, F., \& Singer, S. 1978, Journal of Geophysical Research:
	Space Physics, 83, 75
	
	\bibitem[{Papadopoulos(1985)}]{papadopoulos1985microinstabilities}
	Papadopoulos, K. 1985, Collisionless shocks in the heliosphere: A tutorial
	review, 34, 59
	
	\bibitem[{Perri {et~al.}(2010)Perri, Carbone, \& Veltri}]{perri2010does}
	Perri, S., Carbone, V., \& Veltri, P. 2010, The Astrophysical Journal Letters,
	725, L52
	
	\bibitem[{Raghav {et~al.}(2014)Raghav, Bhaskar, Lotekar, Vichare, \&
		Yadav}]{raghav2014quantitative}
	Raghav, A., Bhaskar, A., Lotekar, A., Vichare, G., \& Yadav, V. 2014, Journal
	of Cosmology and Astroparticle Physics, 2014, 074
	
	\bibitem[{Raghav {et~al.}(2017)Raghav, Shaikh, Bhaskar, Datar, \&
		Vichare}]{raghav2017forbush}
	Raghav, A., Shaikh, Z., Bhaskar, A., Datar, G., \& Vichare, G. 2017, Solar
	Physics, 292, 99
	
	\bibitem[{Raghav \& Kule(2018{\natexlab{a}})}]{raghav2018first}
	Raghav, A.~N., \& Kule, A. 2018{\natexlab{a}}, Monthly Notices of the Royal
	Astronomical Society: Letters, 476, L6.
	\newblock \url{http://dx.doi.org/10.1093/mnrasl/sly020}
	
	\bibitem[{Raghav \& Kule(2018{\natexlab{b}})}]{raghav2018does}
	---. 2018{\natexlab{b}}, Monthly Notices of the Royal Astronomical Society:
	Letters, 480, L6
	
	\bibitem[{Raghav {et~al.}(2018)Raghav, Kule, Bhaskar, Mishra, Vichare, \&
		Surve}]{raghav2018torsional}
	Raghav, A.~N., Kule, A., Bhaskar, A., {et~al.} 2018, The Astrophysical Journal
	
	\bibitem[{Ram{\'\i}rez~V{\'e}lez {et~al.}(2012)Ram{\'\i}rez~V{\'e}lez,
		Blanco-Cano, Aguilar-Rodriguez, Russell, Kajdi{\v{c}}, Jian, \&
		Luhmann}]{ramirez2012whistler}
	Ram{\'\i}rez~V{\'e}lez, J., Blanco-Cano, X., Aguilar-Rodriguez, E., {et~al.}
	2012, Journal of Geophysical Research: Space Physics, 117
	
	\bibitem[{Riley {et~al.}(1995)Riley, Sonett, Balogh, Forsyth, Scime, \&
		Feldman}]{riley1995alfvenic}
	Riley, P., Sonett, C., Balogh, A., {et~al.} 1995, Space Science Reviews, 72,
	197
	
	\bibitem[{Riley {et~al.}(1996)Riley, Sonett, Tsurutani, Balogh, Forsyth, \&
		Hoogeveen}]{riley1996properties}
	Riley, P., Sonett, C., Tsurutani, B., {et~al.} 1996, Journal of Geophysical
	Research: Space Physics, 101, 19987
	
	\bibitem[{Roberts {et~al.}(1987)Roberts, Goldstein, Klein, \&
		Matthaeus}]{roberts1987origin}
	Roberts, D., Goldstein, M., Klein, L., \& Matthaeus, W. 1987, Journal of
	Geophysical Research: Space Physics, 92, 12023
	
	\bibitem[{Roberts {et~al.}(1992)Roberts, Goldstein, Matthaeus, \&
		Ghosh}]{roberts1992velocity}
	Roberts, D.~A., Goldstein, M.~L., Matthaeus, W.~H., \& Ghosh, S. 1992, Journal
	of Geophysical Research: Space Physics, 97, 17115
	
	\bibitem[{Sahraoui {et~al.}(2012)Sahraoui, Belmont, \&
		Goldstein}]{sahraoui2012new}
	Sahraoui, F., Belmont, G., \& Goldstein, M. 2012, The Astrophysical Journal,
	748, 100
	
	\bibitem[{Sahraoui {et~al.}(2006)Sahraoui, Belmont, Rezeau, Cornilleau-Wehrlin,
		Pin{\c{c}}on, \& Balogh}]{sahraoui2006anisotropic}
	Sahraoui, F., Belmont, G., Rezeau, L., {et~al.} 2006, Physical review letters,
	96, 075002
	
	\bibitem[{Sahraoui {et~al.}(2009)Sahraoui, Goldstein, Robert, \&
		Khotyaintsev}]{sahraoui2009evidence}
	Sahraoui, F., Goldstein, M., Robert, P., \& Khotyaintsev, Y.~V. 2009, Physical
	review letters, 102, 231102
	
	\bibitem[{Salem {et~al.}(2012)Salem, Howes, Sundkvist, Bale, Chaston, Chen, \&
		Mozer}]{salem2012identification}
	Salem, C.~S., Howes, G., Sundkvist, D., {et~al.} 2012, The Astrophysical
	Journal Letters, 745, L9
	
	\bibitem[{Schekochihin {et~al.}(2009)Schekochihin, Cowley, Dorland, Hammett,
		Howes, Quataert, \& Tatsuno}]{schekochihin2009astrophysical}
	Schekochihin, A., Cowley, S., Dorland, W., {et~al.} 2009, The Astrophysical
	Journal Supplement Series, 182, 310
	
	\bibitem[{Schrijver \& Siscoe(2010)}]{schrijver2010heliophysics}
	Schrijver, C.~J., \& Siscoe, G.~L. 2010, Heliophysics: Space Storms and
	Radiation: Causes and Effects (Cambridge University Press)
	
	\bibitem[{Schwenn(2006)}]{schwenn2006space}
	Schwenn, R. 2006, Living Reviews in Solar Physics, 3, 2
	
	\bibitem[{Shaikh(2010)}]{shaikh2010inhomogeneous}
	Shaikh, D. 2010, Monthly Notices of the Royal Astronomical Society, 405, 2521
	
	\bibitem[{Shaikh \& Shukla(2009)}]{shaikh20093d}
	Shaikh, D., \& Shukla, P. 2009, Physical review letters, 102, 045004
	
	\bibitem[{Shaikh {et~al.}(2017)Shaikh, Raghav, \& Bhaskar}]{shaikh2017presence}
	Shaikh, Z., Raghav, A., \& Bhaskar, A. 2017, The Astrophysical Journal, 844,
	121
	
	\bibitem[{Smith {et~al.}(2006)Smith, Hamilton, Vasquez, \&
		Leamon}]{smith2006dependence}
	Smith, C.~W., Hamilton, K., Vasquez, B.~J., \& Leamon, R.~J. 2006, The
	Astrophysical Journal Letters, 645, L85
	
	\bibitem[{Sonett \& Abrams(1963)}]{sonett1963distant}
	Sonett, C., \& Abrams, I. 1963, Journal of Geophysical Research, 68, 1233
	
	\bibitem[{Stasiewicz {et~al.}(2000)Stasiewicz, Bellan, Chaston, Kletzing,
		Lysak, Maggs, Pokhotelov, Seyler, Shukla, Stenflo,
		{et~al.}}]{stasiewicz2000small}
	Stasiewicz, K., Bellan, P., Chaston, C., {et~al.} 2000, Space Science Reviews,
	92, 423
	
	\bibitem[{Stawicki {et~al.}(2001)Stawicki, Gary, \& Li}]{stawicki2001solar}
	Stawicki, O., Gary, S.~P., \& Li, H. 2001, Journal of Geophysical Research:
	Space Physics, 106, 8273
	
	\bibitem[{Sundkvist {et~al.}(2007)Sundkvist, Retin{\`o}, Vaivads, \&
		Bale}]{sundkvist2007dissipation}
	Sundkvist, D., Retin{\`o}, A., Vaivads, A., \& Bale, S.~D. 2007, Physical
	review letters, 99, 025004
	
	\bibitem[{Taniuti \& Jeffrey(1964)}]{taniuti1964non}
	Taniuti, A. J.-T., \& Jeffrey, A. 1964, Non-linear wave propagation,  Academic
	Press, New York
	
	\bibitem[{Telloni {et~al.}(2009)Telloni, Bruno, Carbone, Antonucci, \&
		D'Amicis}]{telloni2009statistics}
	Telloni, D., Bruno, R., Carbone, V., Antonucci, E., \& D'Amicis, R. 2009, The
	Astrophysical Journal, 706, 238
	
	\bibitem[{Tsurutani {et~al.}(2011)Tsurutani, Lakhina, Verkhoglyadova, Gonzalez,
		Echer, \& Guarnieri}]{tsurutani2011review}
	Tsurutani, B., Lakhina, G., Verkhoglyadova, O.~P., {et~al.} 2011, Journal of
	Atmospheric and Solar-Terrestrial Physics, 73, 5
	
	\bibitem[{Tsurutani \& Lin(1985)}]{tsurutani1985acceleration}
	Tsurutani, B., \& Lin, R. 1985, Journal of Geophysical Research: Space Physics,
	90, 1
	
	\bibitem[{Tsurutani {et~al.}(1995)Tsurutani, Gonzalez, Gonzalez, Tang, Arballo,
		\& Okada}]{tsurutani1995interplanetary}
	Tsurutani, B.~T., Gonzalez, W.~D., Gonzalez, A.~L., {et~al.} 1995, Journal of
	Geophysical Research: Space Physics, 100, 21717
	
	\bibitem[{Tsurutani {et~al.}(1988)Tsurutani, Gonzalez, Tang, Akasofu, \&
		Smith}]{tsurutani1988origin}
	Tsurutani, B.~T., Gonzalez, W.~D., Tang, F., Akasofu, S.~I., \& Smith, E.~J.
	1988, Journal of Geophysical Research: Space Physics, 93, 8519
	
	\bibitem[{Tsurutani {et~al.}(2018)Tsurutani, Lakhina, Sen, Hellinger,
		Glassmeier, \& Mannucci}]{tsurutani2018review}
	Tsurutani, B.~T., Lakhina, G.~S., Sen, A., {et~al.} 2018, Journal of
	Geophysical Research: Space Physics, 123, 2458
	
	\bibitem[{Verronen {et~al.}(2011)Verronen, Rodger, Clilverd, \&
		Wang}]{verronen2011first}
	Verronen, P.~T., Rodger, C.~J., Clilverd, M.~A., \& Wang, S. 2011, Journal of
	Geophysical Research: Atmospheres, 116
	
	\bibitem[{Vourlidas {et~al.}(2010)Vourlidas, Howard, Esfandiari, Patsourakos,
		Yashiro, \& Michalek}]{vourlidas2010comprehensive}
	Vourlidas, A., Howard, R.~A., Esfandiari, E., {et~al.} 2010, The Astrophysical
	Journal, 722, 1522
	
	\bibitem[{Wal{\'e}n(1944)}]{walen1944theory}
	Wal{\'e}n, C. 1944, Arkiv for Astronomi, 30, 1
	
	\bibitem[{Wu(1987)}]{wu1987mhd}
	Wu, C. 1987, Geophysical Research Letters, 14, 668
	
	\bibitem[{Wu \& Yoon(2007)}]{wu2007proton}
	Wu, C., \& Yoon, P. 2007, Physical review letters, 99, 075001
	
	\bibitem[{Yang \& Chao(2013)}]{yang2013alfven}
	Yang, L., \& Chao, J. 2013, Chin. J. Space Sci, 33, 353
	
	\bibitem[{Yang {et~al.}(2016)Yang, Lee, Chao, Hsieh, Luo, Li, Shi, \&
		Wu}]{yang2016observational}
	Yang, L., Lee, L., Chao, J., {et~al.} 2016, The Astrophysical Journal, 817, 178
	
	\bibitem[{Zhang {et~al.}(2014)Zhang, Moldwin, Steinberg, \&
		Skoug}]{zhang2014alfven}
	Zhang, X.-Y., Moldwin, M., Steinberg, J., \& Skoug, R. 2014, Journal of
	Geophysical Research: Space Physics, 119, 3259
	
	\bibitem[{Zurbuchen \& Richardson(2006)}]{zurbuchen2006situ}
	Zurbuchen, T.~H., \& Richardson, I.~G. 2006, in Coronal Mass Ejections
	(Springer), 31--43
	
\end{thebibliography}

\end{document}